\documentclass{iau_FM}
\usepackage{graphicx}
\usepackage{hyperref}

\title[FM~14.~~Gravitational Wave Symphony of Structure Formation] 
{Asteroseismology of red-giant stars \\as a novel approach in the search for gravitational waves}

\author[Tiago L. Campante et al.]   
{Tiago L. Campante$^{1,2}$
 \and Il\'idio Lopes$^3$
 \and D. Bossini$^{1,2}$
 \and \\ A. Miglio$^{1,2}$
 \and W. J. Chaplin$^{1,2}$}

\affiliation{$^1$School of Physics and Astronomy, University of Birmingham, Edgbaston, \\ Birmingham, B15 2TT, UK \\ email: {\tt campante@bison.ph.bham.ac.uk} \\[\affilskip]
$^2$Stellar Astrophysics Centre (SAC), Department of Physics and Astronomy, \\ Aarhus University, Ny Munkegade 120, DK-8000 Aarhus C, Denmark \\[\affilskip]
$^3$CENTRA, Instituto Superior T\'ecnico, Universidade de Lisboa, Lisbon, Portugal}

\pubyear{2015}
\jname{Astronomy in Focus, Volume~1} 
\editors{Piero Benvenuti, ed.}
\begin{document}

\maketitle

\begin{abstract}
Stars are massive resonators that may be used as gravitational-wave (GW) detectors with isotropic sensitivity. New insights on stellar physics are being made possible by asteroseismology, the study of stars by the observation of their natural oscillations. The continuous monitoring of oscillation modes in stars of different masses and sizes (e.g., as carried out by NASA's {\it Kepler} mission) opens the possibility of surveying the local Universe for GW radiation. Red-giant stars are of particular interest in this regard. Since the mean separation between red giants in open clusters is small (a few light years), this can in principle be used to look for the same GW imprint on the oscillation modes of different stars as a GW propagates across the cluster. Furthermore, the frequency range probed by oscillations in red giants complements the capabilities of the planned {\it eLISA} space interferometer. We propose asteroseismology of red giants as a novel approach in the search for gravitational waves.
\end{abstract}

\firstsection 
\section{Asteroseismology of red-giant stars}\label{sec:astero}
Oscillations in red giants are driven by turbulent motions in the outermost layers of their convective envelopes (for a review, see \cite[Chaplin \& Miglio 2013]{ChaplinMiglio}), this being the same mechanism responsible for the oscillations observed in the Sun and other solar-type stars. But unlike main-sequence stars, red giants have an interior characterized by a high density contrast between core and envelope. This density contrast leads to mixed modes of oscillation that behave as gravity modes in the core and as acoustic modes in the envelope. Over the 4-year nominal mission, {\it Kepler} (\cite[Borucki et al.~2010]{Kepler}) detected oscillations in more than ten thousand red giants up to distances of $15\:{\rm kpc}$, a fraction of which reside in open clusters. This is the case of NGC~6791, which at roughly 8 billion years old (\cite[Brogaard et al.~2012]{cluster}) should be populated by a large number of degenerate stars and hence possible sources of GWs.

\begin{figure}[!t]
\centering
\includegraphics[width=0.9\textwidth]{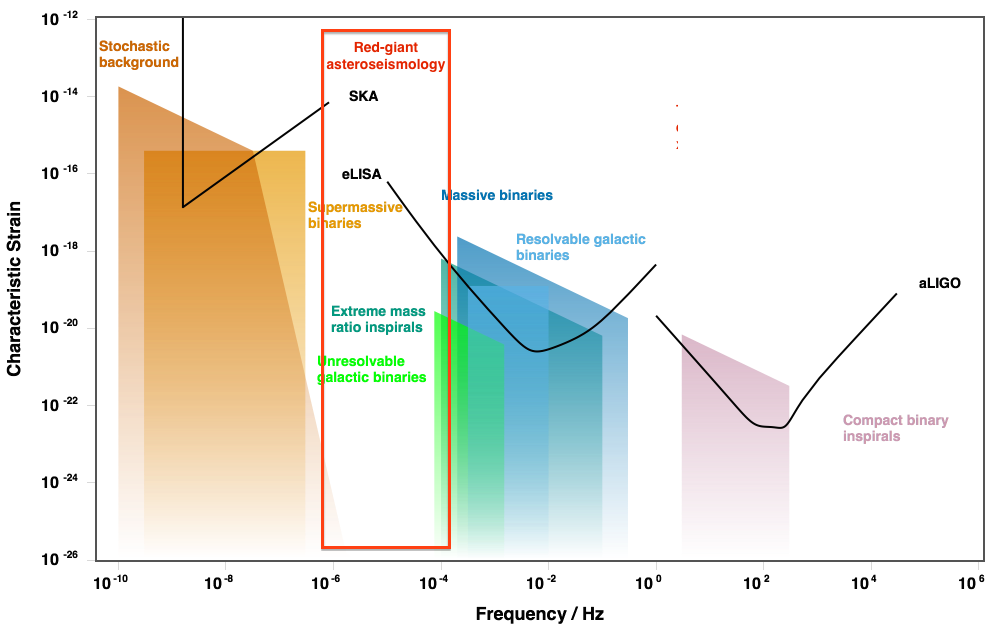} 
\caption{Spectral window probed by red-giant asteroseismology. Also shown are the spectral windows and characteristic strains associated with other gravitational-wave detectors and sources. The authors made use of the Gravitational Wave Sensitivity Curve Plotter available at \url{http://rhcole.com/apps/GWplotter/}.}
\label{fig:strain}
\end{figure}

\section{Red giants as gravitational-wave detectors}\label{sec:gravwaves}
The development of high-precision asteroseismology has opened the possibility of using pulsating stars in the search for GWs (\cite[Lopes \& Silk 2015]{LopesSilk15}). Just like with experimental spherical detectors, the impact of a GW will produce a change in the star's size. As a consequence, one or more modes of oscillation may be affected, resulting in tiny variations of their surface velocity amplitudes. By monitoring such variations it is in principle possible to measure the imprint of a GW on the stellar     frequency-power spectrum. Current models predict that, for the Sun, the surface velocity amplitudes will vary from $10^{-6}$ to $10^{-3}\:{\rm cm\,s^{-1}}$ for GWs with local strains of $10^{-20}$ to $10^{-17}$ (\cite[Lopes \& Silk 2014]{LopesSilk14}), still below current detectability thresholds.

We propose using red-giant stars in stellar clusters to search for GWs. When such stars are located near GW sources (e.g., an ultra-compact binary a few AU away), quadrupole mixed modes may be stimulated above the threshold of detectability. Mode excitation by an external source depends on the internal structure of the star, and in particular on how modes are damped. Mixed modes are characterized by smaller damping rates than those of purely acoustic modes observed in solar-type stars. Moreover, the mean separation between red giants in clusters is small (a few ly). This can be used to look for the same GW imprint on quadrupole modes of different stars. The time lag between the excitation of quadrupole modes of two different stars could be approximately determined from their plane-of-the-sky projected locations and the fact that GWs propagate at the speed of light. Finally, these stars can probe the $10^{-6}$--$10^{-4}\:{\rm Hz}$ spectral window (see Fig.~\ref{fig:strain}), which cannot be fully probed by conventional GW detectors such as the Square Kilometre Array (SKA) or the evolved Laser Interferometer Space Antenna ({\it eLISA}).

%


\end{document}